\documentclass[10pt]{article}
\usepackage{graphicx} % Required for inserting images
\usepackage{placeins}
\usepackage{amsmath}  

\title{\centering Asymmetric binary Bose mixtures: a functional renormalisation group study}
\author{ Boris Krippa\\[1.5em]
\large Department of Mathematics,\\
\large London School of Economics and Political Science, London, UK}
\date{\today}

\begin{document}

\maketitle
\begin{abstract}

We investigate zero-temperature binary Bose mixtures with mass and intraspecies couplings imbalances using the functional renormalisation group (FRG) framework. By analysing the renormalisation-group flow of the condensate densities and the species-dependent wave-function renormalisation factors, we find that this dual asymmetry induces a spontaneous redistribution of quantum fluctuations between the components. This effect leads to a significant divergence in the quasiparticle residues of the two species, a phenomenon not captured in mean-field treatments.  Critically, our FRG analysis predicts a systematic reduction of the density-channel sound velocity with increasing asymmetry—a result that qualitatively contradicts the behaviour predicted by standard Bogoliubov theory. The ratio of the renormalisation factors is proposed as a robust quantitative probe of many-body correlations and fluctuation transfer. These findings provide a consistent non-perturbative framework for interpreting future experiments on heterogeneous mixtures such as  $^{6}\mathrm{Li}\text{-}^{23}\mathrm{Na}$ and $^{41}\mathrm{K}\text{-}^{87}\mathrm{Rb}$.

\end{abstract}

\section{ INTRODUCTION}

Binary Bose mixtures have recently become the subject of intensive theoretical \cite{Pet, Cik, Sta, Ota} and experimental \cite{Gu, Cho, Cab, Fer, Sut, Bur} investigations. In general, quantum mixtures may involve two or more distinct particle species. In the bosonic systems considered in this work, such mixtures can be realised by combining atoms in different spin states or with different masses. They exhibit a rich phase structure, including phase separation \cite{Sut}, droplet formation \cite{Pet}, and superfluid drag \cite{Fil}. Theoretically, these systems have been explored using a variety of approaches, such as Beliaev and Popov theories \cite{Kon, Ute}, Monte-Carlo simulations \cite{Pet1, Cik}, and the functional renormalisation group (FRG) \cite{Isa1}. Despite these efforts, many questions remain open. The central theoretical issue is the interplay between inter- and intraspecies interactions, encoded in the values and signs of the corresponding scattering lengths.

While mean-field (MF) approaches, such as Bogoliubov theory and its extensions \cite{Bo,Sto,Chi}, often provide a reasonable description, they are generally insufficient for capturing the full effect of quantum fluctuations. In particular, it has been shown \cite{Pet} that beyond-mean-field corrections can stabilise mixtures with attractive interspecies interactions that would otherwise collapse within the MF theory. This highlights the need for a nonperturbative framework capable of incorporating fluctuations beyond leading order.

In this paper we use the functional renormalisation group method (FRG) \cite{Wil}. More precisely, we explore the version of FRG \cite{Wet, Mor, Ber, Pa} based on the concept of the average effective action (AEA) $\Gamma_k$. This is a nonperturbative approach where the evolution of the system is represented by the flow in the coupling constant space and depends on some artificial running parameter "k". The corresponding flow is described by the functional flow equation. At some starting scale $k=\Lambda$ the system is represented by some, usually rather simple, classical action and at the end of the evolution at $k=0$ the full quantum action is recovered.

It has been successfully applied across many areas of contemporary theoretical physics, ranging from cosmology and gauge theory \cite{Dup, Reu} to nonrelativistic many-body systems \cite{Kri1, Kri2, Wet1}, few-body physics in atomic and nuclear contexts \cite{Kri2, Die} and one-component Bose-gases \cite{Flo, Bir}.

We note that Monte Carlo (MC) simulations being fully non-perturbative can provide high-precision calculations of the ground-state properties but often struggle with the direct extraction of scale-dependent dynamical factors. Unlike MC the FRG framework offers a transparent view of how mass/coupling constants imbalances drive the system towards or away from the miscibility limit at a non-perturbative level.

The other important feature of FRG is that it provides a natural connection between vacuum and in-medium physics and gives access to the full scale dependence of dynamical quantities, such as wave-function renormalization factors or sound velocities. These factors encode the "dressing" of quasiparticle excitations by quantum fluctuations. In contrast to static thermodynamic observables, these quantities provide direct insight into the low-energy structure of the excitation spectrum and the redistribution of fluctuation weight between components. In the context of Bose mixtures, the behaviour of these factors allows one to probe how the mass  and coupling imbalances fundamentally reshape the dynamical properties of the system.
 While previous FRG studies \cite{Isa1, Isa2} have primarily focused on symmetric mixtures, the present work concentrates on the role of mass and coupling-constant imbalance. We show that these asymmetries lead to a species-dependent renormalisation of quasiparticle residues—an effect that qualitatively modifies the predicted sound velocities and offers a robust probe of many-body correlations.

The paper is organised as follows. In Section 2, we outline the FRG formalism and the specific truncation used. Section 3 discusses the choice of initial conditions. In Section 4, we present our main results, focusing on condensate evolution, stability criteria, and sound velocities. Finally, Section 5 contains our concluding remarks and suggests directions for future research.

\section{FORMALISM}
We use the version of FRG for the AEA $\Gamma_k$ where the dynamics is governed by the following flow equation \cite{Wet,Mor}
\begin{equation}
\partial_k \Gamma_k = - \frac{i}{2} Tr \left [(\partial_k R_k) (\Gamma_{k}^{(2)} - R_k)^{-1}\right].
\end{equation}

Here $\Gamma^{(2)}$ is a second functional derivative of the AEA taken with respect to all fields involved, and the trace $Tr$ includes both loop integration and summation over discrete indices. A cutoff $R_k$ acts as an infrared regulator, which goes to zero at vanishing scale, where the full quantum effective action is recovered and approach $ k^2$ when $k$ is large. 

In spite of seemingly simple one-loop structure this is a functional equation which admits an exact solution in exceptional cases, so in practice one needs to rely on approximations. The most common one is to use the projected flow and construct some symmetry-motivated ansatz for the AEA containing degrees of freedom relevant for relevant energy scale. Using this ansatz one can derive a system of either ordinary or partial differential equations describing the evolution of coupling constants, correlation function etc. with running scale "k".

To analyse the RG flow in the case of a binary boson mixture  we use the following ansatz
\begin{eqnarray}
\Gamma[\psi,\psi^\dagger,\phi,\phi^\dagger,k] = \int d^4x\
[\phi^\dagger(Z_1 (i \partial_t + \mu_1) +\frac{\bar Z_1}{2m_1}\,\nabla^2)\phi  \nonumber\\
+ \psi^\dagger\ (Z_2(i \partial_t + \mu_2)
+\frac{\bar Z_2}{2m_2}\,\nabla^2) \psi 
-U(\rho_1, \rho_2, k)].
\label{eq:ansatz}
\end{eqnarray}
Here $\phi$ and $\psi$ denote two distinctive  bosonic species, $U(\rho_1, \rho_2, k)$ is the scale dependent  effective potential, where we defined $\rho_1=\phi^\dagger\phi$ and $\rho_2=\psi^\dagger\psi$, $Z_{1(2)}$ and $\bar Z_{1(2)}$ are the wave-function and mass renormalisation factors respectively and $m_i(\mu_i)$ are the corresponding masses (chemical potentials). The chemical potentials can be absorbed 
into the potential by redefining
\begin{equation}
    U\rightarrow U - \phi^\dagger Z_1  \mu_1\phi - \psi^\dagger Z_2  \mu_2\psi,
\end{equation}
and the wave-function renormalisation factors $Z_i$ can then be defined as $Z_i = \frac{\partial^2 U}{\partial \mu_i \partial \rho_i}$. The mass renormalisation factors $\bar Z_i $ can in principle be deduced from the effective action but we assume that  $\bar Z_{1(2)}$ do not run with the scale so that their values are taken to be  $\bar Z_{1(2)}$ = 1. Previous analyses \cite{Dup1} indicate that mass renormalisation remains quantitatively weaker than wave-function renormalization in the condensed regime so this assumption seems to be well justified. 

As we have already mentioned the present truncation is expected to be reliable in the dilute regime $n a^3 << 1$. Beyond this regime, taking into account the higher-order terms may become more important and quantitative results should be interpreted with caution.

The classical action  is assumed to take the form

\begin{eqnarray}
S[\psi,\psi^\dagger,\phi,\phi^\dagger,k] = \int d^4x\
[\phi^\dagger((i \partial_t + \mu_1) +\frac{1}{2m_1}\,\nabla^2)\phi  \nonumber\\
+ \psi^\dagger\ ((i \partial_t + \mu_2)
+\frac{1}{2m_2}\,\nabla^2) \psi 
- \sum_i\alpha_i\rho_i - \sum_i\beta_i{\frac{1}{2}}\rho_i^2 \nonumber\\
 - \gamma\rho_1\rho_2]. 
\label{eq:ansatz}
\end{eqnarray}

To obtain the system of the flow equations for the running couplings we expand the effective potential near minima and truncate the expansion at quadratic order
\begin{eqnarray}
 U(\rho_1, \rho_2, k) = U_0 + \sum_i\alpha_i(\rho_i - \rho_{0i}) +  \sum_i\beta_i{\frac{1}{2}}(\rho_i - \rho_{0i})^2 \nonumber\\
 + \gamma(\rho_1 - \rho_{01})(\rho_2 - \rho_{02}).
\end{eqnarray}

We are interested in the condensed phase with a nontrivial vacuum at zero temperature so in what follows we set the coupling $\alpha_i$ to zero. The coupling constants are defined at the minima of the effective potential and run with the scale. The positions of the minima  also run with the scale so we define the total derivative
\begin{equation}   
    d_k = \partial_k + \sum_i  (d_k\rho_{0i}\partial_{\rho_{0i}} + d_k n_{i}\partial_{n_{i}}),  
\end{equation}
where the density $n_i$ is given by the derivative of the effective potential $U$ with respect to chemical potential
\begin{equation}   
    n_i = - \partial_{\mu_i} U |_{(\rho_1\rightarrow \rho_{01},  \rho_2\rightarrow \rho_{02})}.
\end{equation}

We note in passing that the physical density is defined as $n_{ph}(k)=Z(k) \rho_0(k)$.
It is worth mentioning that the RG trajectories may follow different paths depending on what physical parameters are chosen  to run. We consider the RG evolution at fixed chemical potentials, allowing particle densities to run with the scale $k$. Certainly, all possible choices should eventually lead to the same results at the physical point but in practice, due to inevitable truncation of the AEA, it is worth to stick to the trajectories which are better justified by the underlying dynamics of the considered system. In our case the particle densities evolve as a result of interactions, so it seems natural to take this into account even at the intermediate steps.

Applying the total derivative defined above to the couplings, particle densities, renormalisation factors and running minima of the effective potential gives the system of the flow equations (no summation over repeated indices)
\begin{equation}   
   d_k n_i  -  d_{k}\rho_{0i} Z_i = - \partial_{\mu_i} \partial_k U|_{(\rho_1\rightarrow \rho_{01},  \rho_2\rightarrow\rho_{02})},
\end{equation}
\begin{equation}   
    d_k \alpha_i -  d_k\rho_i \beta_i - d_k\rho_j\gamma  =  \partial_{\rho_i} \partial_k U|_{(\rho_1\rightarrow \rho_{01},  \rho_2\rightarrow\rho_{02})}, \qquad (i\neq j)
\end{equation}
\begin{equation}   
    d_k \beta_i  = \partial^2_{\rho_i} \partial_k U|_{(\rho_1\rightarrow \rho_{01},  \rho_2\rightarrow \rho_{02})},
\end{equation}
\begin{equation}   
    d_k \gamma  = \partial^2_{\rho_1,\rho_2} \partial_k U|_{(\rho_1\rightarrow \rho_{01},  \rho_2\rightarrow \rho_{02})},
    \end{equation}
     \begin{equation}   
    d_k Z_i  = -\partial^2_{\rho_i,\mu_i} \partial_k U|_{(\rho_1\rightarrow \rho_{01},  \rho_2\rightarrow \rho_{02})},
\end{equation}
 \begin{equation}   
    d_k U_0  =  \partial_k U|_{(\rho_1\rightarrow \rho_{01},  \rho_2\rightarrow \rho_{02})},
\end{equation}
where some higher-order terms are neglected. This truncation is justified by the dilute nature of the system and by the smallness of the gas parameter $\sqrt{n a^3} <<1$, ensuring the dominance of leading-order derivative terms.  A possible and probably better alternative would be to estimate the higher order terms using a mean-field (MF) approximation. This option is left for future studies. We kept the coupling $\alpha_i$ in the flow equations to take into account the possibility of a symmetry restoration at some intermediate scale.

The driving terms (right hand sides of the equations 8-13) can be obtained by substituting the ansatz for the AEA and performing a contour integration. The poles in $k\rightarrow 0$ limit correspond to typical Bogoliubov's spectra for the Bose mixture \cite{Pet1} which, in the case of the symmetric mixture ($m_1=m_2=m, \beta_1=\beta_2=\beta$ and $\rho_1=\rho_2=\rho)$ at the physical point $k = 0$ can be written as

 \begin{equation}   
    E_{\pm} =\sqrt {\frac{q^2}{2 m}(\frac{q^2}{2 m} + 2\rho(\beta\pm\gamma))}.
\end{equation}

The important ingredient of the FRG technique is the cutoff function $R_k$. We use the optimised cutoff version suggested in \cite{Lit}.

\begin{equation}   
    R_{k,i}(q) = \frac{1}{2 m_i}(k^2 - q^2)\Theta(k^2 - q^2),
\end{equation}
where $\Theta(x)$ is the step function.
This cutoff significantly simplifies the calculations of momentum integrals and optimises the RG path in a sense defined in \cite{Lit}. 
We note that, while a variety of regulator functions are available in the FRG framework, we have employed the  regulator, which is the standard choice for stability and convergence in the derivative expansion and which is known to be optimal for the truncation level used here. We expect the observed qualitative effects—specifically the species dependent dynamical renormalisation to be robust features of the imbalanced system rather than regulator-dependent artifacts.

On the other hand, this cut-off may lead to irregularities when calculating the higher order correlation functions or dealing with the mixed-boson-fermion systems, where this cutoff function does not provide any simplifications when calculating the boson-fermion loop integrals. In these cases, the use of smooth cutoffs looks preferable. 

\section{INITIAL CONDITIONS}

To extract initial conditions (IC) we follow the strategy outlined  in \cite{Isa1}. The FRG flow starts at some initial UV scale $k =\Lambda$  where $\Gamma_{k=\Lambda}$ is given by the classical action $S$  so the IC can be determined from the requirement $\Gamma_{k=\Lambda} = S$. Then it follows that $\rho_{i,0}(\Lambda) = n_{i}(\Lambda)$  and we consider the densities $n_i$ at starting scale as input.

For the renormalisation factors $Z_i$ we take $Z_{1,2}(\Lambda) = 1$ and, as we always work in the condensed phase we assume $\alpha_i=0$.  The IC for the couplings $\gamma, \beta_i$ can be determined as follows \cite{Isa1, Wet1}. The starting scale is supposed to be large compared to the typical many-body scales of the system which are commonly set by the chemical potentials. Therefore in order to fix the values of these couplings at $k =\Lambda$ it is reasonable to analyse an auxiliary flow, starting from vacuum ($\rho_{i,0}=\mu_i=0)$ with $k=0$ and up to $k=\Lambda$. At $k=0$ the couplings $\beta_i, \gamma$ are related to the scattering lengths which are observables and can at least in principle  be extracted from the experimental data. The repulsive (attractive) nature of the boson-boson interactions will be reflected in the signs of the scattering lengths and will eventually show up in the resulting phase diagram of the many-body system. To extract the starting values of the chemical potentials we used the mean field type of expression $\mu_i=\beta_i n_i + \gamma n_j$.

Following this line of reasoning, one can extract the IC for the couplings $\gamma, \beta_i$ to be

\begin{equation}
    \beta_i(\Lambda)=\frac{4\pi/m_i}{a_{i}^{-1} - 4\Lambda/3 \pi} ,
\end{equation}
and similarly for the interspecies coupling $\gamma$

\begin{equation}
    \gamma(\Lambda)=\frac{2\pi/m_{ij}}{a_{ij}^{-1} - 4\Lambda/3 \pi}  ,
\end{equation}
where $m_{ij} = m_i m_j/(m_i + m_j)$ is  the corresponding reduced mass.

It is worth mentioning that, as noted in \cite{Wet1}, the microscopic scale $\Lambda^{-1}$ must not be significantly larger than the scattering length for the pointlike interaction to remain a valid approximation. Moreover, for the cutoffs taken to be too large, the theory becomes "trivial" so that in what follows, the constraint $\Lambda \leq 1/a$ is always imposed. The initial values of the densities can be obtained using identities $n_i(\Lambda) = \rho_i(\Lambda)$

The stability of the potential at the negative interspecies interaction requires $\beta_1\geq 0$, $\beta_2\geq 0$ and $\beta_1\beta_2\geq\gamma^2$. One notes that interspecies coupling can also be positive, which implies that the corresponding interaction is repulsive, but in the following we will be interested in the case where the coupling $\gamma$ takes negative values.

\section{RESULTS}
In our numerical analysis, we set the UV cutoff to unity ($\Lambda=1$).  In these units, the evolution of the system is initiated at a density $n/\Lambda^3=0.1$. This choice reflects the dilute gas regime, where the characteristic gas parameter 
 remains small $(\sqrt{n a^3}<<1)$. All other dimensionful quantities are also expressed in units of the ultraviolet scale 
$\Lambda$ according to their canonical dimensions.

We find that the RG flow remains numerically stable throughout the entire integration range down to the infrared (IR) limit. This choice of parameters allows for a clear manifestation of the fluctuation-induced effects—specifically the species-dependent dynamical renormalisation—while ensuring that the results are not obscured by numerical instabilities or artifacts of the truncation.

We begin by showing in Fig.1 the behaviour of the condensates as a function of the mass ratio $R = m_1/m_2$ at fixed couplings $\beta_1=2.1, \beta_2=2.2$ and $\gamma=-0.1$. We consider a range of asymmetries from $R=1$ to $R=4$ that would roughly correspond to physical systems such as $^{87}\mathrm{Rb}\text{-}^{85}\mathrm{Rb}$ ($R \simeq 1$) and $^{6}\mathrm{Li}\text{-}^{23}\mathrm{Na}$ ($R \simeq 4$).

\begin{figure}[h!]
  \centering
  \includegraphics[width=0.9\linewidth]{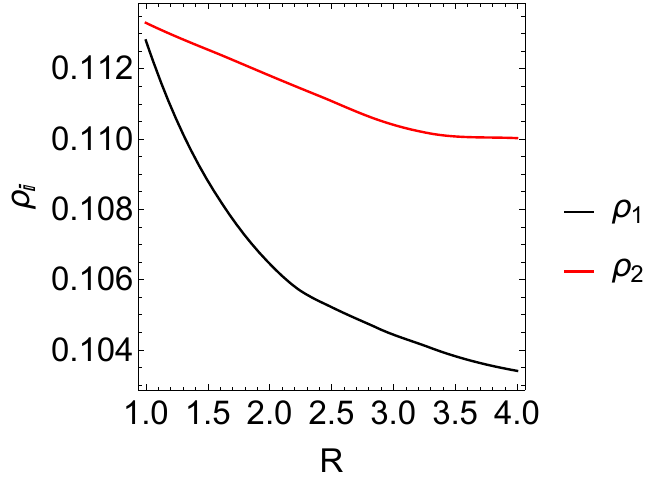}
  \caption{Two condensates $\rho_1$ (lower curve, black online) and $\rho_2$ (upper curve, red online) as functions of the mass ratio $R$. The interspecies coupling is $\gamma = -0.1$}
\end{figure}
As seen in Fig.~1, both condensates decrease with increasing mass asymmetry, although the effect is relatively weak. In contrast, when the mass asymmetry is reduced, the two condensates approach each other. The remaining finite difference between them near $R\simeq 1$  arises from the difference in the intraspecies coupling constants. When these couplings are taken to be equal and the interspecies coupling is set to zero, the condensates coincide.

Physically, these results imply that when the mass asymmetry increases the quantum fluctuations of the lighter particle become more prominent. Moreover, fluctuations result in increase of the effective repulsion leading thus to the shift of the minimum of the thermodynamical potential towards the smaller densities. From the formal side the lighter component of the mixture dominates in the loop contributions effectively leading to suppression of the condensates.
It is worth noting that all this is a pure beyond-mean-field effect, which is elusive in the MF computations.

\begin{figure}[h!]
  \centering
  \includegraphics[width=0.9\linewidth]{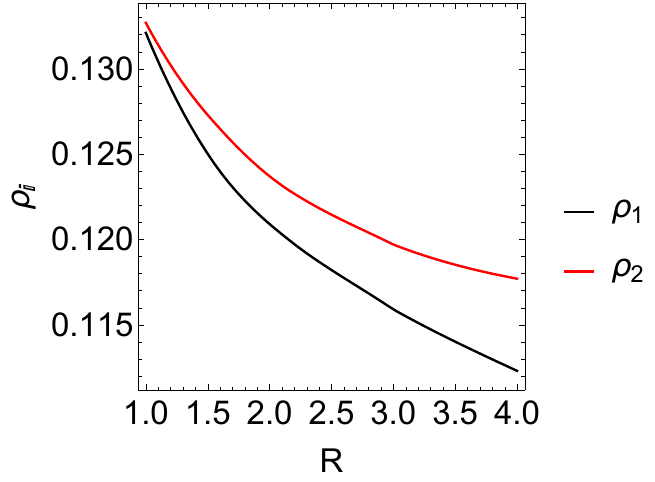}
  \caption{Same as for Fig.1 but with interspecies coupling  taken at the initial scale to be $\gamma=-1 $}
\end{figure}

As follows from Figs. 1-2 both condensates increase when the attraction between different species becomes larger and effectively deepens the effective potential. In other words the increased interaction strength between different bosonic species further compresses the mixture. However, when the  coupling $\gamma$ approach the critical value $\gamma^2\simeq\beta_1 \beta_2$ the system becomes unstable and eventually collapses. Physically it corresponds to the fact that the interaction matrix stops being positive definite so that the Gross-Pitaevsky equation no longer has a stationary solution.

\begin{figure}[h!]
  \centering
  \includegraphics[width=0.9\linewidth]{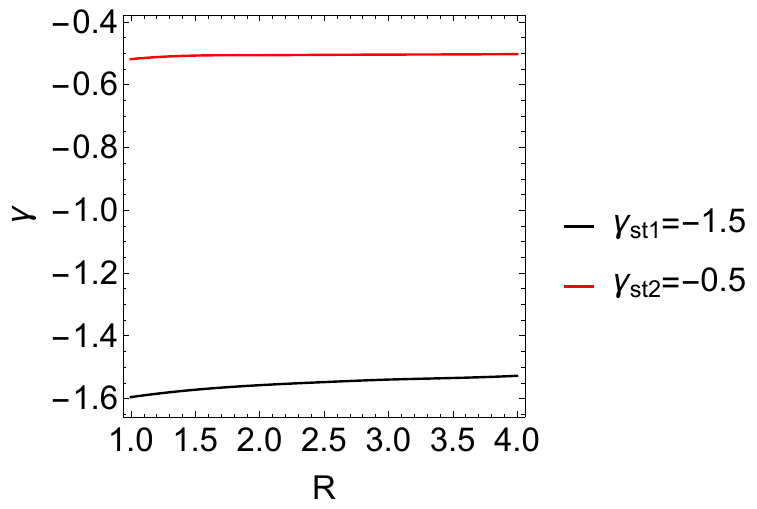}
  \caption{The interspecies coupling $\gamma$ as a function of the mass ratio $R$. The starting values of the couplings are  $\gamma=-0.5$ and $\gamma=-1.5$ (red and black online)} 
  
\end{figure}

In Fig. 3 we show the dependence of the interspecies coupling $\gamma$ on the mass ratio $R$. In Fig.3 the starting values of the couplings are taken to be $\gamma_{st}(\Lambda)= -0.5 (-1.5)$. As one can see from Fig.3 the physical values of the couplings slightly increase  as the mass asymmetry becomes larger. It implies that the interaction effectively becomes less attractive as the RG scale $k$ approaches physical point $k = 0$ making the system of two distinct bosonic species with a small/moderate mass asymmetries more stable with respect to a possible collapse. When the starting value of the  coupling is taken to be $\gamma(\Lambda)= -1.5$  this tendency becomes more pronounced. This behaviour reflects the effect of  the  fluctuations which can  partly compensate an interspecies attraction. When the  coupling increases the system strives to form strongly coupled correlated state eventually leading to collapse. 

In Figs. 4-5 we show the dependence of the condensates on the interspecies coupling $\gamma$ for asymmetries $R=1$ and $R=3$. 
\begin{figure}[h!]
  \centering
  \includegraphics[width=0.9\linewidth]{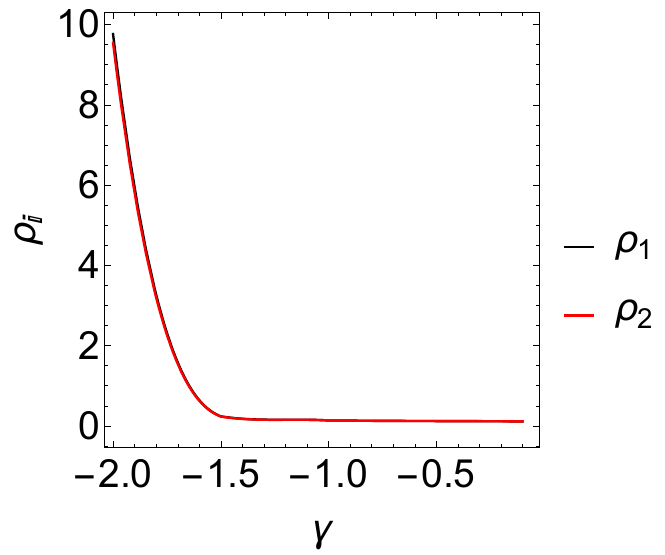}
  \caption{The condensates as the functions of the starting value of the  coupling  $\gamma$ at $R=1$}
\end{figure}

\begin{figure}[h!]
  \centering
  \includegraphics[width=0.9\linewidth]{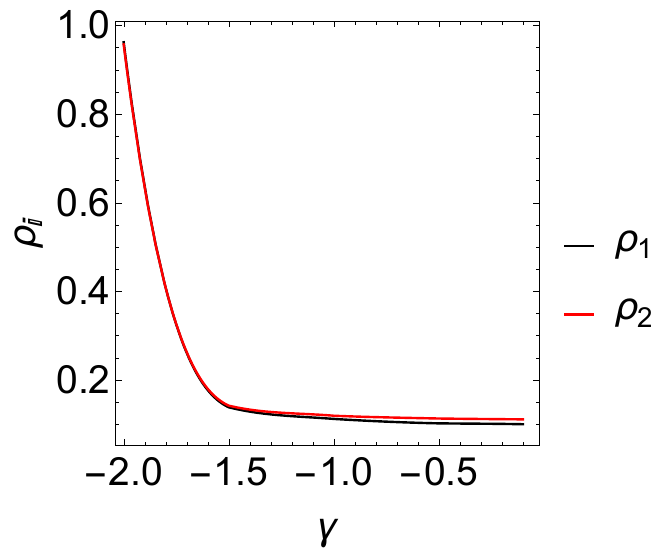}
  \caption{Same as in Fig.4 but for the asymmetry factor $R=3$.}
\end{figure}

\begin{figure}[h!]
  \centering
  \includegraphics[width=0.9\linewidth]{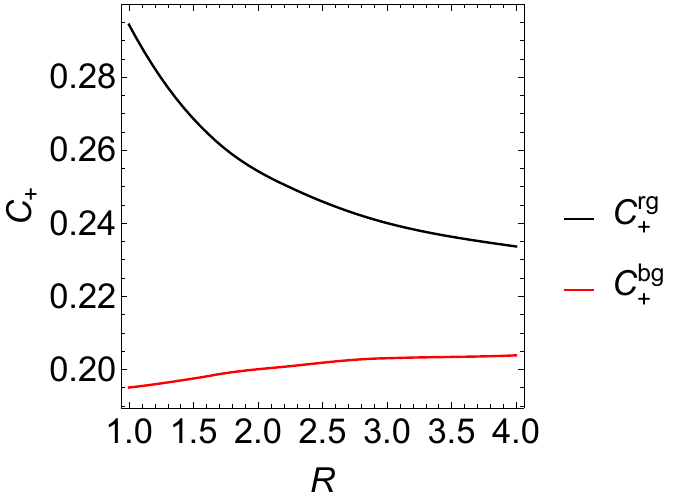}
  \caption{Sound velocity in the density channel calculated in FRG (black online) and in the Bogoliubov approximation (red online) as the function of $R$. The starting value of the interpecies coupling is taken to be  $\gamma_{st}=-1$}
\end{figure}

One can conclude that the increase of the mass asymmetry does not really affect significantly this dependence in the region of small/moderate$ \gamma$. In both cases the condensates start rising sharply starting at $\gamma \simeq -1.5$ when the attraction generated by this coupling becomes comparable with the repulsion between the same bosonic species and dominate over kinetic pressure. At larger couplings the growth of the condensates becomes even faster eventually leading to collapse. The effect is stronger for the symmetric ($R=1$) case.
  
 \begin{figure}[h!]
  \centering
  \includegraphics[width=0.9\linewidth]{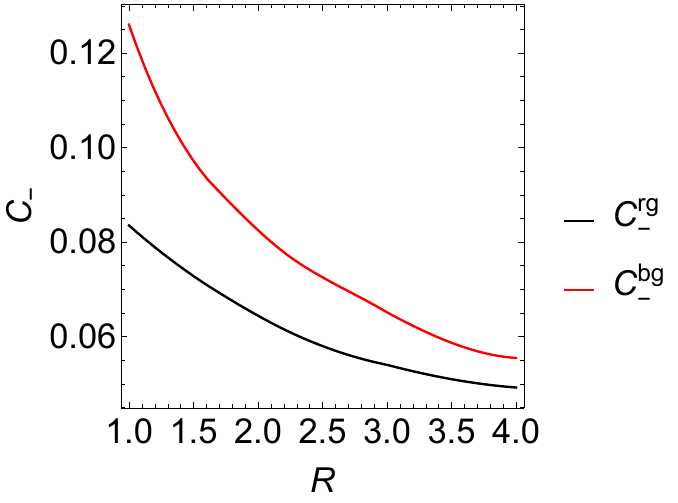}
  \caption{Same as in Fig.6 but for the sound velocity in the spin channel.}
\end{figure}

An important characteristic of the Bose-Bose mixtures is provided by the sound velocities. It is  well known that in this system the low-energy excitation spectrum consists of two sound modes namely density ($c_+$) and spin ($c_{-}$) ones which are shown in Figs. 6-7 calculated in the FRG and Bogoliubov approximation. An interesting qualitative difference can be seen from Fig.6 between the Bogoliubov and the FRG results for the behavior of the density mode velocity 
 as a function of asymmetry. While within the Bogoliubov approximation $c_{+}$ exhibits a slight increase with increasing asymmetry, the RG flow instead leads to a systematic reduction of this velocity.

The increase of $c_+$ at the mean-field level can be understood as follows. As asymmetry between the components grows, the contribution of the stiffer component may dominate, leading to a modest enhancement of the overall compressibility modulus and, consequently, of the sound velocity in the density channel.

In contrast, the RG treatment incorporates fluctuation effects that qualitatively modify this picture. Fluctuations lead to a scale-dependent renormalization of both the interaction parameters and the wavefunction renormalization factors $Z_i$. In particular, they tend to soften the effective interactions and increase the compressibility of the system, resulting in a reduction of the density sound velocity.

The results for the sound velocity in the spin channel are shown in Fig.7. Both within the Bogoliubov approximation and in the FRG framework, we find that the spin velocity decreases as the asymmetry increases.

This behavior has a clear physical origin. The spin mode describes relative motion of the two fluids; its stiffness is therefore controlled by how strongly the components resist being displaced with respect to each other. In a symmetric system, both components respond similarly to perturbations, leading to a relatively stiff mode and a larger value of $c_{-}.$ However, once asymmetry is introduced, one component becomes effectively “softer” and can more easily adjust to the motion of the other. As a result, relative density fluctuations cost less energy, leading to a reduction of the spin sound velocity.

This effect is already captured at the Bogoliubov level, where $c_{-}$ is determined by the eigenvalues of the interaction matrix. However, the RG treatment enhances this tendency. Fluctuations lead to a scale-dependent renormalization of the couplings and wavefunction factors, typically amplifying the imbalance between the components and further softening the spin channel.  

We numerically verified that in the weak-coupling limit, the FRG results smoothly approach those obtained within Bogoliubov theory, providing a useful consistency check for the present numerical scheme. As the interaction strength increases, the sound velocities predicted by Bogoliubov theory become complex already before the bare classical stability threshold is reached. In this intermediate regime, the FRG approach still yields real-valued sound velocities, indicating that fluctuation effects can substantially delay or soften the apparent mean-field (Bogoliubov) instability.

 We show in Fig.8 the behavior of the ratio of the respective renormalisation factors $Z_2/Z_1 - 1$ as a function of the interspecies coupling constant. Physically, the wave-function renormalization factor determines the quasiparticle residue and thus directly controls the spectral weight of elementary excitations. Deviations of $Z_i$ from unity reflect the degree of many-body dressing and provide a sensitive measure of interaction-induced dynamical correlations beyond the mean-field level.
 \begin{figure}[h!]
  \centering
  \includegraphics[width=0.8\linewidth]{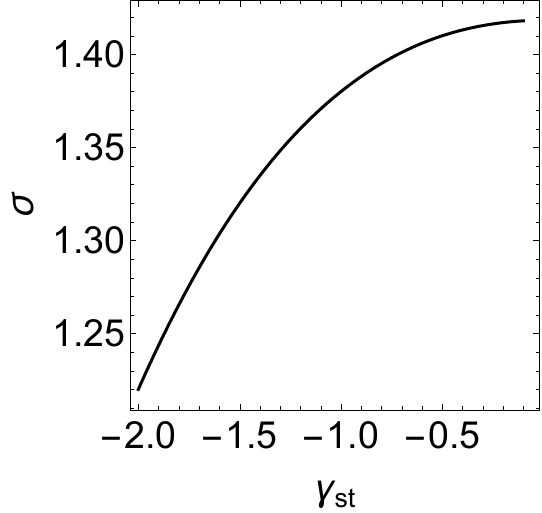}
  \caption{The value of $\sigma=(Z_2/Z_1 - 1)$ scaled by the factor $10^2$ versus  the interspecies coupling at initial scale $\gamma_{st}$ at the mass asymmetry $R=4$.}
\end{figure}

 Although the overall effect is numerically rather small it represents a purely dynamical many-body contribution, absent at the mean-field level. Its smallness reflects the weakly interacting character of dilute Bose mixtures, while its systematic dependence on mass imbalance and interspecies coupling reveals subtle collective effects inaccessible within simpler theoretical frameworks like mean-field approach. Mass imbalance leads to different wave-function renormalizations  which modifies the effective superfluid stiffness of each component and therefore affects the velocities of the density and spin collective modes.

We observe from Fig.8 that  the quantity $Z_2/Z_1 - 1$ decreases monotonically with increasing  the  attraction between species, even for strongly asymmetric systems.

This behavior indicates that the dynamical properties of the two components become progressively more similar under the RG flow. Physically, this can be understood as a consequence of the attractive coupling, which promotes correlated fluctuations between the components. As the inter-species interaction strengthens, the system tends to minimize its energy by aligning the fluctuations of the two condensates, effectively locking their dynamics.

From the RG perspective, this effect originates from loop contributions involving mixed propagators, which couple the flow of renormalisation factors.
 As a result, the initially distinct wavefunction renormalisations are driven toward each other. In the strong-coupling regime, the system behaves increasingly like a single effective fluid, characterized by nearly identical dynamical coefficient.

 We find that the wave-function renormalisation of the heavier component of the mixture is slightly more pronounced than that of the lighter one, and the difference increases with mass asymmetry. This behavior can be attributed to the softer low-energy excitation spectrum of the heavier species. In other words, quantum fluctuations more efficiently dress the heavy component, leading to a stronger suppression of its quasiparticle residue and fluctuation redistribution in the Bose mixtures.

Overall, the analysis of the wave-function renormalization factors uncovers subtle dynamical effects arising from the interplay of mass imbalance and interspecies interactions, and demonstrates the capability of the FRG approach to resolve delicate many-body phenomena beyond the reach of mean-field and perturbative methods.

One notes that, at the fixed interspecies coupling $\gamma = -0.13$ and varying mass ratio, the relative dynamical renormalization decreases almost monotonically as the mass ratio approaches unity, dropping by almost 6 times when going from R = 4 to R = 1. This behavior reflects the progressive restoration of dynamical symmetry between the components and provides a direct measure of the mass-induced  quantum fluctuations.

\begin{figure}[h!]
  \centering
  \includegraphics[width=0.9\linewidth]{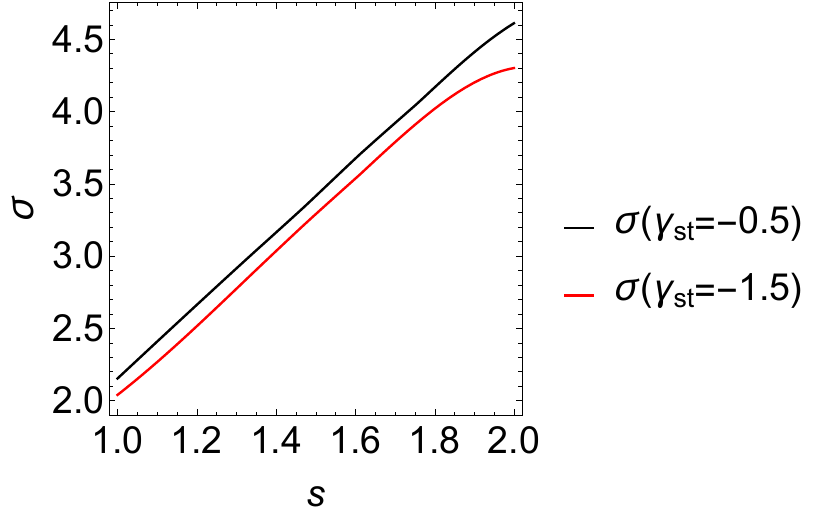}
  \caption{The value of  $\sigma=(Z_2/Z_1 - 1)$ scaled by the factor $10^2$  versus the factor $s=\beta_2/\beta_1$ at starting values of interspecies couplings taken to be $\gamma_{st}=-0.5$ (red online) and $\gamma_{st}=-1.5$ (black online)}. 
\end{figure}

Next we analysed the behaviour of the renormalisation factors as the function of the intraspecies couplings. The results are shown in Fig.9. The dynamical renormalisation factor $\sigma$  at $s=\beta_2/\beta_1=2$  is found to be between $4\%$ and $4.5\%$. As one can see this factor scales approximately linearly with  $s$ in the region $s\simeq 1$. This behavior demonstrates that in this region the relative stiffness of the two components is predominantly controlled by their respective interaction strengths. 
The dependence on the interspecies coupling  $\gamma$ is found to be weak and provides only  subleading corrections in the regime considered. In particular, varying 
$\gamma$ by a factor of three results only in a relatively minor upward shift $\sim 5\%$ of the corresponding curve, without altering its qualitative behavior.  We emphasize, however, that this hierarchy is not universal and may change in other regions of parameter space.  In the regime of moderate asymmetry this linear scaling breaks down for larger values of s
 where clear deviations from linearity are observed (already at $s\simeq 2$ where the difference becomes $\sim 10\%$. This indicates a crossover from a symmetric regime, where each component is predominantly renormalized by its own interaction, to a strongly asymmetric regime where nonlinear effects and intercomponent coupling become increasingly important.

\begin{figure}[h!]
  \centering
  \includegraphics[width=0.9\linewidth]{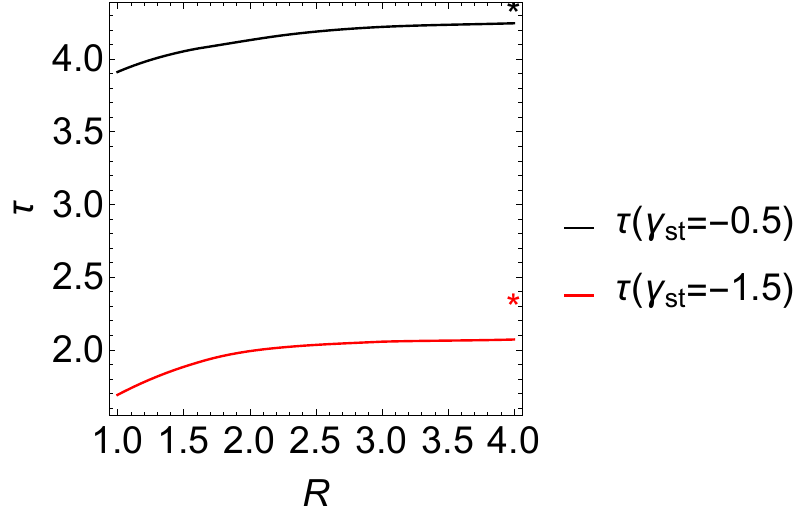}
  \caption{The stability factor $\tau$ calculated as a function of $R$ at starting values of interspecies couplings taken to be $\gamma_{st}=-0.5$ (black online) and $\gamma_{st}=-1.5$ (red online). The starting values of the stability factors are $4.38 $ and $2.36$ respectively (the stars on the graph)}
\end{figure}

Fig.10 shows  the FRG results for the stability factor defined as $\tau=\sqrt{\beta _1\beta_2 -\gamma^2}$ for two starting values of the interpspecies couplings. The stability factors decrease during the RG flow which implies that at the vanishing scale the system is less stable than at the beginning of the evolution. It suggests that the system may collapse even if the bare classical stability condition is still satisfied.

The stability factor increases with growing asymmetry between the components, approaching  (but never reach) its initial (ultraviolet) value  defined at the beginning of the RG flow. At $R=4$ this tendency is particularly pronounced for smaller initial values of the inter-species coupling, where the rate of increase is  higher.

This behavior can be understood as a consequence of the reduced efficiency of inter-species correlations in asymmetric systems. As the asymmetry (e.g., in masses or intraspecies interactions) increases, the two components respond differently to fluctuations, which weakens their mutual coupling at low energies. As a result, the system becomes effectively less correlated and moves away from the instability region.

The dependence on the initial inter-species coupling further supports this interpretation. For weaker initial attraction, fluctuation effects are less efficient in driving the system toward instability, and the decoupling induced by asymmetry dominates more rapidly. This leads to a faster recovery of the stability factor. In contrast, for stronger initial attraction, inter-species correlations remain significant over a larger range of scales, slowing down this restoration.

\FloatBarrier
\section{DISCUSSION AND CONCLUSION}

We have studied binary boson mixtures at zero temperature with mass and coupling constants asymmetries in the framework of the functional renormalisation group (FRG). While previous FRG studies have successfully mapped the phase diagrams of symmetric Bose mixtures, the role of mass imbalance in shaping the dynamical dressing of quasiparticles has remained largely unexplored. This work provides non-perturbative evidence that mass asymmetry induces a spontaneous redistribution of quantum fluctuations, leading to a noticeable divergence in the quasiparticle residues of the two species.

We demonstrated that this relative dynamical renormalisation arises solely due to the combined effect of interspecies interactions and asymmetry in the masses and intraspecies couplings. In the absence of interspecies coupling and with equal masses, the two components renormalise independently, resulting in identical wave-function renormalisation factors. In other words, mass and coupling constants imbalances generate species-dependent dynamical renormalisation, which grows with increasing attraction between species and serves as a quantitative indicator of fluctuation redistribution between components. Crucially, we demonstrate that this effect results in a systematic reduction of the density-channel sound velocity—a prediction that qualitatively contradicts the behaviour observed in standard Bogoliubov theory. By identifying these species-dependent renormalisation factors as a robust probe of many-body correlations, we offer a quantitative framework for future experiments on  $\text{Li}\text{-}^{23}\text{Na}$ and $^{41}\text{K}\text{-}^{87}\text{Rb}$ .

Regarding the sound velocity in the spin channel, we found that within both the Bogoliubov approximation and the FRG framework, it decreases as mass asymmetry increases. Fluctuations, which lead to a scale-dependent renormalisation of the couplings and wave-function renormalisation factors, typically amplify the imbalance between the components, further softening the spin channel.

It is interesting that as the interspecies coupling increases the dynamical instability in the spin channel occurs before the bare classical stability criterion is formally violated. At these couplings, the spin-wave velocity becomes imaginary even while the effective potential still supports a stable local minimum. This indicates that quantum fluctuations trigger a collapse via the spin sector before the classical phase separation or collapse threshold are reached. 
 
 We find that the behaviour of the collective modes near the instability threshold depends qualitatively on the mass ratio. For equal masses, the density and spin sound velocities become nearly degenerate as the system approaches the instability, indicating an emergent symmetry between the modes. In contrast, for large mass asymmetry, this degeneracy is lifted and a clear separation between components persists even close to the instability. This demonstrates that mass imbalance in some sense stabilises the distinction between density and spin channels, preventing their complete hybridisation.

We have also analysed the behaviour of the condensates and the interspecies coupling. We find that decreasing mass asymmetry enhances the condensate values and drives them closer to each other, while increasing interspecies attraction leads to rapid condensate growth as the system approaches the mean-field instability threshold.
 Beyond this point, the present truncation indicates instability and the RG flow cannot be continued. The RG evolution of the stability criterion reveals that quantum fluctuations tend to reduce the stability region compared to the mean-field prediction. In particular, we observe that a system that is stable at the bare level may flow towards a less stable regime as quantum fluctuations are integrated out towards the infrared  limit indicating that the standard miscibility condition overestimates stability. This effect is most pronounced at weak interspecies coupling and becomes less significant at stronger coupling where the bare classical stability condition remains effectively valid.
 
We emphasise that, within the present truncation, the flow does not explicitly resolve the formation of self-bound droplets beyond the mean-field instability. Specifically, the non-analytic Lee–Huang–Yang contribution is not captured explicitly. Nevertheless, the FRG approach provides a consistent nonperturbative description of the approach to instability, including fluctuation-induced renormalisation of dynamical observables and collective modes. A full description of the droplet phase would require an extended truncation that includes higher-order terms in the effective potential or explicitly accounts for beyond-mean-field stabilisation mechanisms.

Overall, our nonperturbative analysis demonstrates the crucial role of quantum fluctuations in stabilising ultracold Bose mixtures and reveals subtle dynamical effects related to the wave function renormalisation factors which are inaccessible within mean-field approaches. Crucially, this species-dependent renormalisation acts as a feedback loop: the redistribution of fluctuation energy modifies the effective interspecies interaction, which in turn leads to a 'softening' of the density-channel sound velocity. Unlike the standard Bogoliubov approach, where the quasiparticle residues are fixed, the FRG framework demonstrates that mass asymmetry inherently destabilises the mixture more rapidly than mean-field theory predicts. This explains why the system reaches the miscibility-immiscibility transition earlier in the RG flow, providing a dynamical explanation for the shift in the stability phase boundary. These effects should be experimentally testable in ultracold atomic systems with tunable mass imbalance and interaction strengths, opening a direct route to probing dynamical renormalisation phenomena in multicomponent quantum fluids.

 It is important to note that the dynamical redistribution of fluctuations is driven by a dual asymmetry: the mass imbalance and the disparity in the intraspecies couplings. We demonstrate that even if the  interaction between bosons of the different type is weak, the combined effect of these two asymmetries forces the system into a regime where the two components experience rather different polarisation. This leads to the 'relative' dynamical renormalisation that serves as the hallmark of our work, providing a more complex and realistic picture than the standard symmetric or purely mass-imbalanced cases."        

Several extensions of the present work are worth pursuing. In particular, incorporating higher-order terms in the effective potential may allow one to describe fluctuation-stabilised droplet phases. Further generalisations include mixtures with more than two components, with obvious applications to QCD at finite density where the competition between chiral, meson, and colour condensates is expected \cite{Kle,Kli}. Taking into account finite temperature effects, where additional collective phenomena are expected, is another possible direction for future studies.

  \section{Acknowledgments}
  The author would like to thank A. Rubtsov for a number of useful discussions.


\begin{thebibliography}{99}

\bibitem{Pet}  D. S. Petrov, Phys. Rev. Lett., \textbf{115}, 155302, (2015).

\bibitem{Cik} V. Cikijevic et al,  Phys. Rev. A \textbf{99}, 023618, (2019).

\bibitem{Sta} C. Staudinger, F. Mazzani and R. E. Zillich, Phys. Rev. A \textbf{98}, 023633, (2018).

\bibitem{Ota}  M. Ota and G. E. Astrakharchik, SciPost Phys., \textbf{9}, 020, (2020). 

\bibitem{Gu}  Q. Gu and L. Yin, Phys. Rev. B \textbf{102}, 220503, (2020). 

\bibitem{Cho} L. Chomas et al, Phys. Rev. X \textbf{ 6}, 041039, (2016).

\bibitem{Cab} C. Cabrera et al, Science \textbf{359}, 301, (2018).

\bibitem{Fer} G. Ferioli, Phys. Rev. Letters \textbf{122},  090401, (2019).

\bibitem{Sut}  K. Suthar, A. Roy, D. Angom, Phys. Rev. A \textbf{91}, 043615, (2015). 

\bibitem{Bur}  A. Burchianti et al, Condens. Matter \textbf{5}, 21, (2020). 

\bibitem{Guo}  Z. Guo et al, Phys. Rev. Res \textbf{3}, 033247 (2021). 

\bibitem{Fil} D. V. Fil, S. I. Shevchenko, Phys. Rev. A \textbf{72}, 013616, (2005).

\bibitem{Kon} P. Konietin and V. Pastukhov, Journal of Low Temperature Physics, \textbf{190}, 256, (2018).

\bibitem{Ute} O. I. Utesov, M. I Baglay and S. V. Andreev, Phys. Rev. A \textbf{97} 053617, (2018).

\bibitem{Pet1} D. S. Petrov and G. E. Astrakharchik, Phys. Rev. Letters, \textbf{117}, 100401, (2016).

\bibitem{Isa1} F. Isaule  et al, Phys. Rev. A \textbf {104}, 023317, (2021).

\bibitem{Dup1} N. Dupuis, Phys. Rev. E \textbf {83}, 031120, (2011).

\bibitem{Bo} N. N. Bogoliubov,  J. Phys. (Moscow), \textbf {11}, 23, (1947).

\bibitem{Sto} J. Armaitis, H. T. C. Stoof, and R. A. Duine,  Phys. Rev. A \textbf{91},  034641, (2015).

\bibitem{Chi} E. Chiquillo, Phys. Rev. A \textbf { 97}, 063605, (2018).

\bibitem{Wil} K. G. Wilson, Rev. Mod. Phys. \textbf {55}, 583, (1983). 

\bibitem{Wet} C. Wetterich, Phys. Lett. B \textbf {301}, 90, (1993).

\bibitem{Mor} T. Morris, Int. J. Mod. Phys. A \textbf { 9}, 2411, (1994).

\bibitem{Ber} J. Berges, N. Tetradis and C. Wetterich, Phys. Reports \textbf {363}, 223, (2002).

\bibitem{Pa}  J. M. Pawlowski, Annals Phys. A \textbf {322}, 2831, (2007).

\bibitem{Dup} N. Dupuis et al,  Phys. Reports \textbf {910}, 1, (2021).

\bibitem{Reu} M. Reuter, F. Saueressig, Phys. Rev. D \textbf {65}, 065016, (2002). 

\bibitem{Kri1}  B. Krippa,  J. Phys. A \textbf{42}, 465002, (2009). 

\bibitem{Wet1} I. Boettcher, J. M. Pawlowski, and C. Wetterich, Phys. Rev. A \textbf{89}, 053630, (2014).  

\bibitem{Kri2} M. C. Birse, B. Krippa and N. R. Walet, Phys.  Rev.  A \textbf, {83}, 023621, (2011).

\bibitem{Die} R. Schmidt and S. Moroz, Phys. Rev. A \textbf{81}, 052709, (2010).

\bibitem{Flo} S. Floerchinger and C. Wetterich, Phys. Rev. A  \textbf {77}, 053603, (2008).

\bibitem{Bir} F. Isaule, M. C. Birse and N. R. Walet, Annals Phys.  \textbf {412}, 168006, (2020).

\bibitem{Isa2} F. Isaule and I. Morera, Cond. Matter  \textbf{7}, 9, (2022).

\bibitem{Lit} D. F. Litim, Journal of High Energy Physics,  \textbf{2001}, 059, (2001).

\bibitem{Gri} Y.Shi and A. Griffin,  Phys. Rep.  \textbf{304}, 1, (1998). 

\bibitem{Liu} H. Hu, J. Wang and X. Liu,  Phys. Rev. A \textbf{102}, 043301, (2020). 

\bibitem{Kle} B. K. Klein, D. Toublan and J. J. M Verbaarshot and X. Liu,  Phys. Rev. D \textbf{72}, 015007, (2005).

\bibitem{Kli} T. G. Khunjua, R. G. Klimenko  and R. N. Zhokhov, Progress in Theoretical and Experimental Physics,  \textbf{2}, 023B03, (2025). 


  















\end{thebibliography}
\end{document}